\documentclass[final,3p,times,twocolumn]{elsarticle}
 \biboptions{comma,sort&compress}
\usepackage{here}
 \usepackage{graphicx}
  \usepackage{epsfig}
\usepackage[]{lineno}

\def\nin{\noindent}
\def\beq{\begin{equation}}
\def\eeq{\end{equation}}
\def\bea{\begin{eqnarray}}
\def\eea{\end{eqnarray}}

\journal{Nuc. Phys. (Proc. Suppl.)}

\begin{document}

\begin{frontmatter}



\title{Precision measurements with jets and particles at HERA}

 \author[label1]{Katharina M\"uller\corref{cor1} \\ on behalf of the H1 and ZEUS Collaborations }

  \address[label1]{Physikinstitut der Universit\"at Z\"urich, 
Winterthurerstrasse 190,
CH - 8057 Z\"urich, Switzerland.}
\cortext[cor1]{Speaker}
\ead{kmueller@physik.uzh.ch}


\begin{abstract}
\noindent
Inclusive jet and multijet cross sections measured in $ep$ collisions in different kinematic regions by the H1 and ZEUS
experiments are shown.
The measurements are used to extract the strong coupling  $\alpha_s$ as a function of the scale and at the mass of the $Z$-boson.
Results from prompt photon production in comparison with pQCD calculations
are shown in photoproduction and deep-inelastic scattering.
Other topics address  scaled momentum distributions of charged particles,
 a measurement of the transverse momenta of charged particles and the first measurement of the charge asymmetry at HERA.

\end{abstract}


\end{frontmatter}


\section{Introduction}
\nin
The HERA $ep$-collider operated with electrons or positrons of energy  27.6~GeV and protons of 920~GeV. Each of the two experiments H1 and ZEUS collected roughly 0.5~fb$^{-1}$ in fifteen years of running. 
At HERA, two kinematic regimes are distinguished, deep-inelastic scattering (DIS) with high  photon virtualities, $Q^2\!\ge\! 5$~GeV$^2$, and photoproduction ($\gamma p$) with a quasi real photon, $Q^2\!\simeq\! 0$~GeV$^2$, which is the dominating process.  
The large statistics of the HERA data allows detailed tests of perturbative QCD (pQCD) based on  cross section measurements with jets or prompt photons. 
In DIS there are two relevant hard scales, $Q$ and the transverse momentum $P_T$ of the jet or the photon, while in $\gamma p$ there is only $P_T$.
Jet measurements in DIS are carried out in the Breit frame.  The $k_T$ cluster algorithm, which is collinear and     infrared safe, is used to reconstruct  jets.
\nin
\section{Jet production}
Jet production in $ep$ scattering provides stringent tests of QCD and an independent assessment of the gluon contribution to the parton density functions (PDFs). It further allows the extraction of the strong coupling  $\alpha_s$ as a function of the scale and at the mass of the $Z$-boson.

In five recent analyses jet production was measured at HERA in various kinematic regions.
 ZEUS measured the inclusive jet cross section in $\gamma p$~\cite{km_zeusjet1} ($Q^2\!<\!1$~GeV$^2$, $P_{T,jet}\!>\!17$~GeV
) and high $Q^2$ DIS~\cite{km_zeusjet2} ($Q^2\!>\!125$~GeV$^2$, $P_{T,jet}\!>8$~GeV). Both analyses provide a determination of $\alpha_s$, furthermore the 2-jet cross section is measured for the high $Q^2$~\cite{km_zeusdijet} region ($125\!<\!Q^2\!<\!20000$~GeV$^2$, $P_{T,jet}\!>\!8$~GeV). Here the invariant mass of the two jets is required to be $M_{j,j}>\!20$~GeV. 
The 2-jet cross section has a high sensitivity to the gluon contribution of the PDFs in kinematic regions where its uncertainty is contributing significantly to the theoretical error.  H1 measured inclusive, 2-jet and 3-jet cross sections as well as the ratio of the 2-jet to 3-jet cross sections for low~\cite{:2009he} ($5\!<\!Q^2<100$~GeV$^2$, $5<\!P_{T,jet}\!<80$~GeV, $M_{j,j}\!>\!18$~GeV) and high $Q^2$~\cite{Aaron:2009vs} ($150\!<\!Q^2\!<\!15000$~GeV$^2$, $7\!<\!P_{T,jet}\!<50$~GeV, $M_{j,j}\!>\!16$~GeV). For the latter the jet cross sections are normalised to the inclusive DIS cross section, which significantly reduces the experimental and theoretical errors. 

\begin{figure}[hbt] \hspace*{-0.5cm}
\includegraphics[width=5cm]{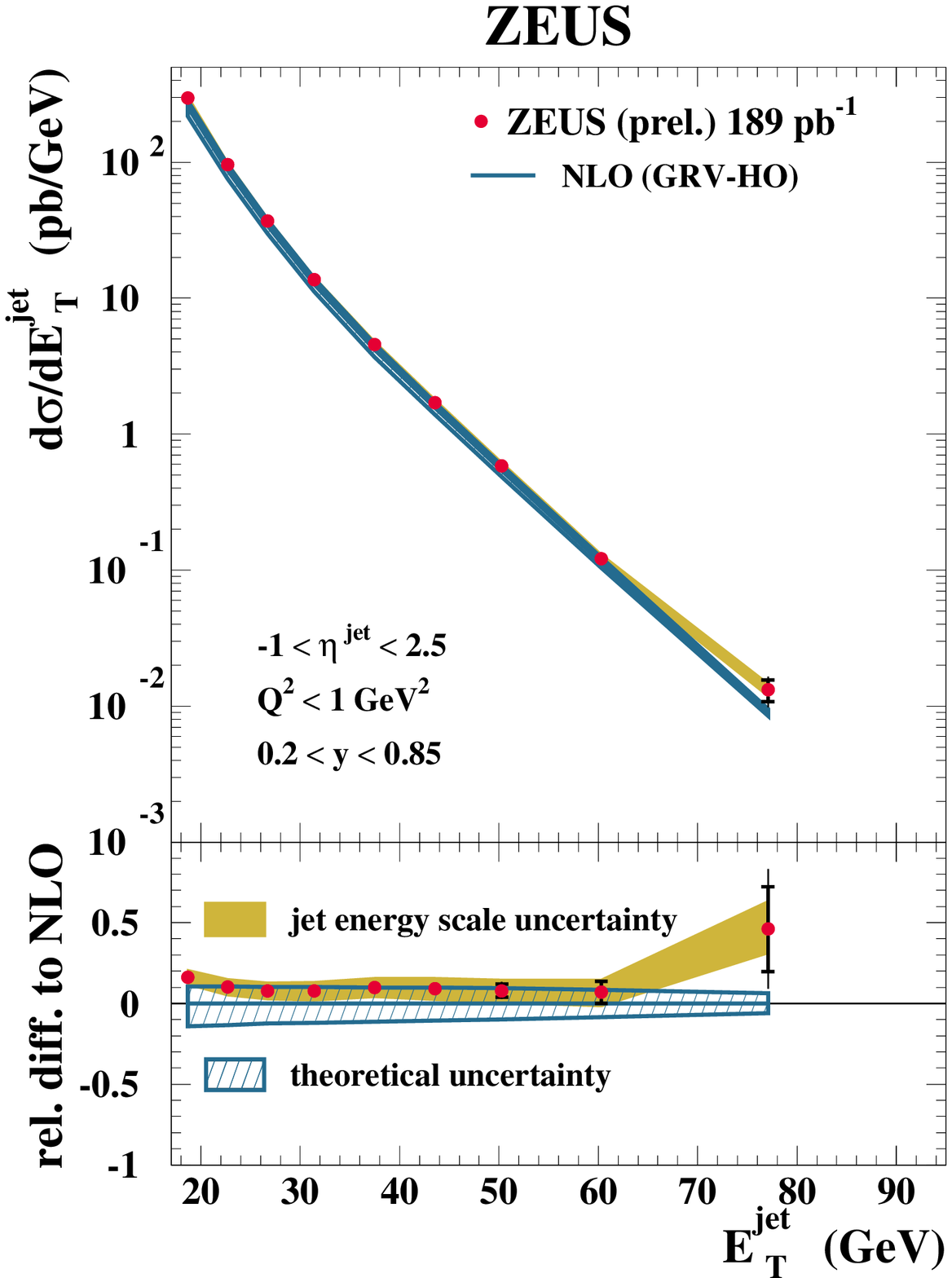}\hspace*{-1cm}
\includegraphics[width=5cm]{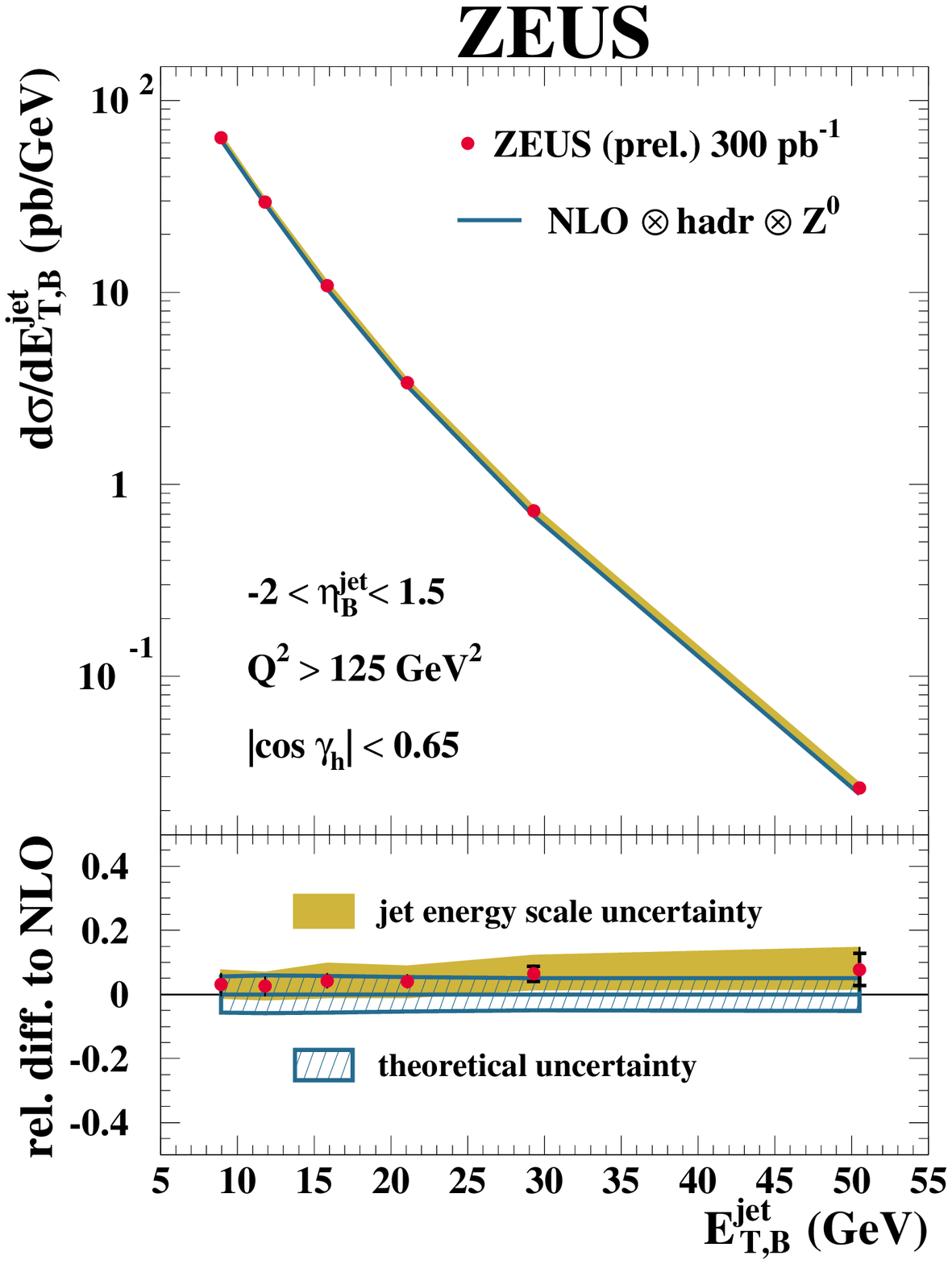}
\caption{\scriptsize Differential cross section for inclusive jet production in photoproduction (left) with $Q^2\!<\!1$~GeV$^2$ and NC DIS (right) with ($125\!<\!Q^2\!<\!20000$~GeV$^2$). Jets are found with the longitudinally inclusive $k_T$-algorithm  in the Breit frame. }
\label{km_zeusjet} 
\end{figure}

Fig.~\ref{km_zeusjet} shows the inclusive jet cross section as a function of the transverse energy of the jet for the ZEUS measurement in $\gamma p$ and high $Q^2$ DIS.
For both kinematic regions the cross section falls steeply. The data are very precise, the dominant experimental error being the uncertainty of the energy scale  of the jets, which is $1(3) \%$ for jet energies above (below) $10$~GeV.
The measurement is compared to  NLO QCD predictions which describe the measurement very well. The theoretical and experimental errors are of comparable size. The theoretical errors are dominated  by the renormalisation scale uncertainty.

Similar results are obtained for multijet cross sections. Overall the description by the NLO calculations is very good in all kinematic regions, though the low $Q^2$ analysis suffers from large theoretical uncertainties of up to $30\%$.
\subsection{Extraction of running $\alpha_s$ and $\alpha_s(M_Z)$}
\begin{figure}[hbt] 
\centerline{\includegraphics[width=5.cm]{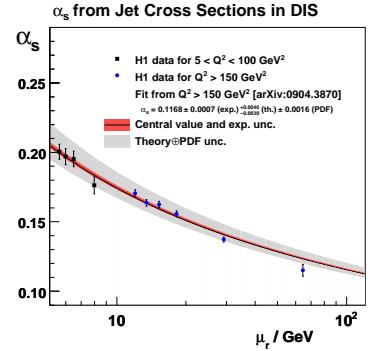}}
\caption{\scriptsize $\alpha_s$ as a function of the scale $\mu_r\!=\!\sqrt{(Q^2+P_{T,jet}^2)/2}$ from jet cross sections at low and high $Q^2$.}
\label{km_h1run} 
\end{figure}

\nin
The jet cross sections discussed above are used to extract the strong coupling $\alpha_s$ at different values of the renormalisation scale $\mu_r$ and at the $Z$-boson mass. Statistical, systematic and correlated uncertainties are taken into account. The dominant theory uncertainty is estimated by varying the renormalisation and factorisation scales  by a factor 0.5 and 2.
The running of $\alpha_s$ as a function of the renormalisation scale is shown in Fig.~\ref{km_h1run}. The values are  extracted from the H1 data at low and high $Q^2$, similar results exist from ZEUS. In the low $Q^2$ region the values and experimental uncertainties are found to be in  good agreement with the QCD expectation which is based on the extracted value of $\alpha_s(M_Z)$ of the high $Q^2$ measurement.  

\begin{figure}[hbt] 
\centerline{\includegraphics[width=6.5cm]{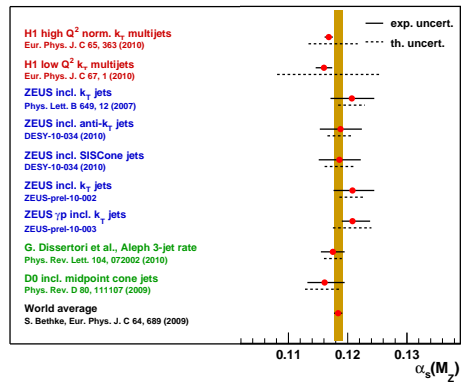}}
\caption{\scriptsize Recent values of $\alpha_s$ from HERA, LEP and TEVATRON.}
\label{km_sumalphas} 
\end{figure}

Fig.~\ref{km_sumalphas} shows a summary of recent $\alpha_s$ measurements using jets at HERA together with the most precise determination of $\alpha_s$ from LEP~\cite{Dissertori:2009qa} and  from TEVATRON~\cite{Abazov:2009nc}. 
All measurements are in  good agreement with each other in different kinematic regions ($\gamma p$, low and high $Q^2$) and with the world average.
 Further measurements  using different jet algorithms (anti-$k_T$ and SISCone~\cite{Abramowicz:2010ke}) lead to similar results. 
For many of the precise HERA results the theoretical uncertainties dominate the error. Higher order calculations are expected to improve the results.
\nin

\section{Prompt Photon Production}
\begin{figure}[hbt]  
\centerline{\includegraphics[width=6cm]{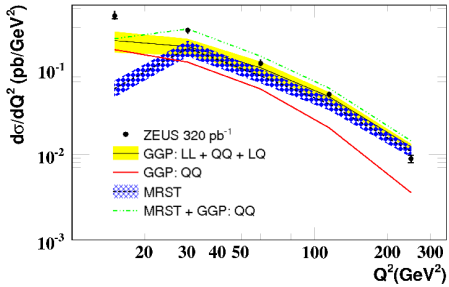}}
\caption{\scriptsize Isolated photon differential cross section $d\sigma/dQ^2$. The data is compared to a order $\alpha^3$ calculation (GGP) as well as to a calculation with an increased contribution of the LL part (MRST).}
\label{km_promptg} 
\end{figure}

\nin
Events with an isolated photon emerging from the hard subprocess $ep \rightarrow e\gamma X$
- so called prompt photons -  offer an alternative method to study  hard interactions.
Recent results have been published on prompt photon production in $\gamma p$~\cite{Aaron:2010uj} for transverse energies of the photon $6\!<\!E_T^\gamma\!<\!15$~GeV  by H1 and in DIS~\cite{Collaboration:2009dqa} ($4\!<\!E_T^\gamma\!<\!15$~GeV) by ZEUS. Both experiments use the shower shapes of the electromagnetic cluster to discriminate the signal of single photons from multiple photons of decays of neutral hadrons. As already observed in previous publications, the new results show that the available calculations are not able to  describe all the measured distributions well. 
In photoproduction it is found that the NLO calculations  underestimate the inclusive prompt photon cross section, while there is reasonable agreement for events with a prompt photon and a jet. 
However, their transverse correlation, which is sensitive to higher order processes, is not well described.
In DIS the cross section receives contributions from radiation off the electron (LL) and also off the quark (QQ).
The differential  cross section as a function of $Q^2$ is shown in  Fig.~\ref{km_promptg}. The order $\alpha^3$ QCD prediction (GGP)  significantly underestimates the data at low $Q^2$. 
Also included in the figure is the prediction of MRST for the LL part of the cross section. This calculation is based on QED contributions to the PDFs which increases the LL contribution. MRST together with the QQ contribution from GGP  shows similar deficits at low $Q^2$.

\section{Charged Particle Production}
\subsection{Rapidity spectra of charged particles}
\begin{figure}[hbt] 
\centerline{\includegraphics[width=5.5cm]{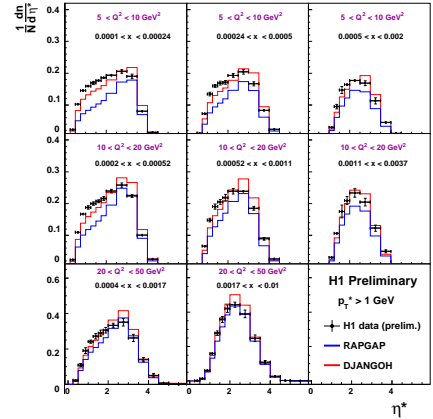}}
\caption{\scriptsize Rapidity spectra of charged particle with $p_T^\star\!>\!1$~GeV for different bins of $Q^2$ and $x$.}
\label{km_parton_shower} 
\end{figure}

\nin
HERA experiments are able to access very small Bjorken $x$, a kinematic region where it is expected that the parton dynamics differs from the description by the DGLAP evolution equations. The latter imply  a strong ordering of the transverse momenta $k_T$ in the parton cascade from the proton to the virtual photon. 
Measurements of the hadronic final state are sensitive to the dynamics of the parton cascade.
Fig.~\ref{km_parton_shower} shows the $\eta^*$ spectra of charged particles with a transverse momentum in the 
hadronic centre of mass system  $p_T^*\!>\!1$~GeV in different bins of $x$ and $Q^2$ as measured by H1 in low $Q^2$ DIS events~\cite{km_H1_transverse}. 
The data are compared to two MC predictions. 
RAPGAP is based on the DGLAP evolution equations for the parton dynamics, whereas DJANGOH  follows the
Color Dipole Model, in which parton radiation is not ordered in $p_T$.
At small $x$ and $Q^2$ and in the forward (proton) direction the RAPGAP predictions  are significantly below the data, whereas  the data are described reasonably well over the full kinematic range by the approach based on the
Color Dipole Model.
\subsection{Scaled momentum distributions}
\begin{figure}[hbt] 
\centerline{\includegraphics[width=5.5cm]{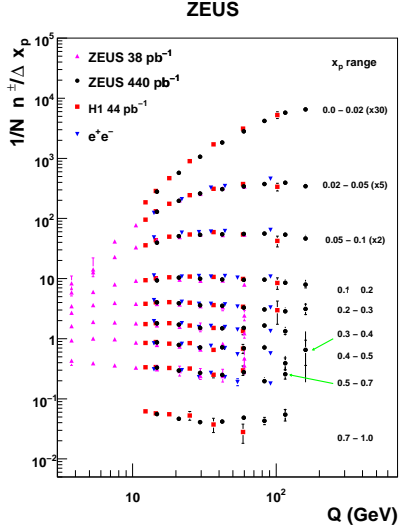}}
\caption{\scriptsize $n^*$ distribution in the hadronic centre of mass system of charged particles with $p_T^*\!>\!1$~GeV.}
\label{km_scaling} 
\end{figure}

\nin
Quark fragmentation may be studied at HERA by using the scaled momentum $x_p$ in the current region of the Breit frame as observable. 
Here, $x_p\!=\! 2P_{Breit}/Q$ with $P_{Breit}$ the momentum of a hadron. 
Scaled momentum distributions were measured by ZEUS for  $10\!<\!Q^2\!<\!41000$~GeV$^2$ for tracks with a transverse momentum larger than 0.15~GeV.
Fig.~\ref{km_scaling} shows the density of charged particles per unit of $x_p$ as a function of $Q$ in bins of $x_p$. As the energy scale $Q$ increases, the phase space for soft gluon emission increases, leading to a rise of the number of particles with small $x_p$, which is clearly seen in Fig.~\ref{km_scaling}. In this figure the data from HERA are shown together with data from $e^+e^-$ which were scaled to half of the centre-of-mass energy. The overall agreement between the different data sets supports the concept of fragmentation universality. 
 NLO calculations predict  too weak scaling violation and also Monte Carlo predictions are not able to describe the data in the full kinematic range.

\subsection{Hadronic charge asymmetry}
\begin{figure}[hbt]
\centerline{\includegraphics[width=5.5cm]{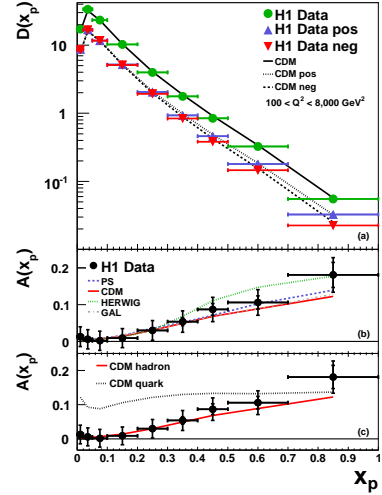}}
\caption{\scriptsize Normalised distribution of the scaled momentum $D\!=\!1/N dn/dx_p$ (a) for all, positively charged and negatively charged particles and charge asymmetry $A(x_p)$ as a function of $x_p$. The data is compared to Monte Carlo predictions with different parton cascade and hadronisation processes and to the parton level before the hadronisation.}
\label{km_asym} 
\end{figure}

\nin
The hadronic charge asymmetry were measured in the Breit frame by H1~\cite{:2009vh}.
Fig.~\ref{km_asym} shows the event normalised distribution of the scaled momentum for all, positively and negatively charged particles. There are significantly more particles produced at low $x_p$ than at high $x_p$. 
At low $x_p$ the distribution is very similar for negative and positive particles. This is expected since low $x_p$ particles are predominantly produced in fragmentation. Fig.~\ref{km_asym}c illustrates that the original asymmetry observed on quark level is not visible anymore on hadron level at low $x_p$. At high $x_p$ there is an overshoot of positively charged particles reflecting the charge asymmetry of the proton.
The asymmetry is reproduced by various models. The data are expected to further constrain the valence quark distributions in the proton and to provide useful information on the fragmentation functions. 

\section{Conclusions}
\nin
Several results of inclusive jets and multijets production in different kinematic regions have been presented. 
The accurate measurements are well described by NLO calculations and allow the extraction of the strong coupling  $\alpha_s$ as a function of the scale and at $M_Z$ with 
small experimental errors.

Results of prompt photon production in photoproduction and DIS are compared
to theoretical predictions. They  have problems describing the data in some kinematic regions both in $\gamma p$ and DIS. In general they underestimate the data, most significantly at low $Q^2$ in DIS.

Measurements of the hadronic final states are used to study the parton dynamics and fragmentation processes.
The charged particle spectra at low $Q^2$ and low $x$ are sensitive to the parton dynamics and comparisons to models hint at dynamics beyond the conventional DGLAP evolution equations at NLO. 
The scaled momentum spectra have been measured in DIS. 
Large scaling violations are observed. Comparing the data to $e^+e^-$ results supports the concept of quark-fragmentation universality.
The hadronic charge asymmetry was measured and found to be largest at large scaled momenta $x_p$. 
The results are consistent with the expectation that at high $x_p$ the asymmetry is directly related to the quark content of the proton.


\section*{Acknowledgements}
\nin
I like to thank the organisers of this interesting conference and my colleagues from H1 and ZEUS for valuable inputs to the talk and these proceedings.







\end{document}